\documentclass[preprint,showpacs,showkeys,groupedaddress,superscriptaddress]{revtex4}
\usepackage{epsf,epsfig}
\usepackage[psamsfonts]{amssymb}
\usepackage{amsmath}
\usepackage{bm}
\usepackage{natbib}

\begin{document}

\title{Casimir effects of objects in fluctuating scalar and electromagnetic fields: Thermodynamic investigatinge}

\author{M. Jafari  }
\email{jafary.marjan@gmail.com\\ m.jafari@sci.ikiu.ac.ir} \affiliation{Department of Physics, Faculty of Science, Imam Khomeini International University, 34148 - 96818, Gazvin, Iran}

\date{\today}

\begin{abstract}
Casimir entropy is an important aspect of casimir effect. In this paper, we employ the path integral method  to obtain a general relation for casimir entropy and internal energy of arbitrary shaped objects in the presence of two, three and four dimension scalar fields and the electromagnetic field. We obtain the casimir entropy and internal energy of two nano ribbons immersed in the scalar field and two nanospheres immersed in the scalar field and the electromagnetic field. The casimir entropy of two nanospheres immersed in the electromagnetic field behaves differently in small interval of temperature variations. 
\end{abstract}


\maketitle
\section{Introduction}\label{Introduction}
\noindent  Since the zero point energy of vacuum fluctuations plays a very important role in the theory of quantized fields, it has been introduced into various branches of physics \cite{2-1}.\\
  The casimir effect occurs due to the boundary conditions of medium surfaces imposed on the fluctuating field. In the other words, the casimir energy refers to the difference between the energy of the fluctuating field when objects are presence and when the objects are removed to infinity \cite{1-3,12-3}, therefore, the casimir entropy has been extensively studied \cite{4,5,6,7}.\\
  Recently, a group of scientists presented a multi-scattering formalism to study fluctuating fields and casimir effects \cite{2-2,3-2,4-2}. They used multi-scattering formalism for bodies when they were weakly coupled to the quantum fields \cite{16-8,111,114}.\\
  In 2012, we used path integral techniques to calculate the casimir force of arbitrary shaped objects immersed in a fluctuating field. It was shown that, in the first order approximation in our method might, the casimir energy be equivalent to weak coupling limit in the multiple-scattering formalism \cite{8}.\\
   Recently, scientists have found that the intervals of negative entropy also occur for some geometries \cite{2,3,7-2,8-2,4,5}. Since the specific heat of system is proportional to the variation of the system entropy, where coupling between system and reservoir can be negligible, \cite{2,5,2-5}, the specific heat of system is defined by the difference between the specific heat of the system as well as reservoir and specific heat of the reservoir. Therefore, in cases where two parallel plates \cite{7-2} and a sphere beside a plate are analysed according to Drude model \cite{8-2}, casimir entropy becomes negative.\\
    In this paper, we investigate casimir entropy, using the path integral methods for arbitrary shaped objects which are immersed in a massless scalar field and an electromagnetic field \cite{8}.
 First, we obtain a general relation for casimir entropy and internal energy of the system. Then, we apply this procedure to objects immersed in (1+1)D, (2+1)D, (3+1)D scalar fields, and the electromagnetic field.\\ 
    This paper is organized as follows: in section II, we briefly describe the free energy and entropy based on path integral methods; in section III, we calculate them for 1+1 Dimension, 2+1 Dimension and 3+1 Dimension space-time in the scalar field; Finally in section IV, we obtain free energy and entropy in the electromagnetic field. 
    
\section{Free energy and entropy }\label{Free energy and entropy}
\noindent   The casimir free energy of a system in a quantum field theory at finite temperature in general form becomes as follows [appendix A]
\begin{equation}\label{2ee}
E=-\textit{k}_B T \sum _{l=0} ^\infty tr \ln[G(i\nu_l;\textbf{x},\textbf{x}')],
\end{equation}
 where, $G(i\nu_l;\textbf{x},\textbf{x}')$ is total Green's function of interacting system which defines
\begin{eqnarray}\label{expansion Green function}
G(\textbf{x}-\textbf{x}',\omega)=G^0(\textbf{x}-\textbf{x}',\omega)+ \int_\Omega d^n \textbf{z}_1 G^0(x-z_1,\omega)
[\omega^2 \tilde{\chi}(\omega,\textbf{z}_1)]G^0(\textbf{z}_1-\textbf{x}',\omega)+ \nonumber\\
\int_\Omega \int_\Omega d^n \textbf{z}_1 d^n \textbf{z}_2
G^0(\textbf{x}-\textbf{z}_1,\omega) [\omega^2
\tilde{\chi}(\omega,\textbf{z}_1)]G^0(\textbf{z}_1-\textbf{z}_2,\omega)
[\omega^2
\tilde{\chi}(\omega,\textbf{z}_2)]G^0(\textbf{z}_2-\textbf{x}',\omega)
+\cdots,
\end{eqnarray}
where $\tilde{\chi}(\omega,\textbf{x})$ is the susceptibility
function of the medium with frequency variable and $G^0(\textbf{x}-\textbf{x}',\omega)$ is Green's function of the medium.
In terms of the susceptibility function, the expansion of free energy is achieved by
\begin{eqnarray}\label{exp1}
E&=&\textit{k}_B T \sum_{l=0}^\infty \sum_{n=1}^\infty
\frac{(-1)^{n+1}}{n}\int d^n\textbf{x}_1\cdots d^n\textbf{x}_n G^0
(i\nu_l;\textbf{x}_1-\textbf{x}_2)...G^0(i\nu_l;\textbf{x}_n-\textbf{x}_1) \nonumber \\
&\times& \chi(i\nu_l,\textbf{x}_1)...\chi(i\nu_l,\textbf{x}_n),
\end{eqnarray}
in which $G_0(i\nu,\textbf{x}-\textbf{x}')$ is green's function of the system without reservoir.\\
 The casimir entropy is obtained from the casimir free energy, using thermodynamic relation
\begin{equation}\label{2}
S =- \frac{{\partial E}}{{\partial T}}
\end{equation}
According to the Eqs.(\ref{2ee}) and (\ref{2}), in terms of total Green's function, casimir entropy of the system is 
\begin{equation}
S = K_B \sum\limits_{l = 0}^\infty  {tr} [LnG(i\upsilon _l ;x,x') + TG^{ - 1} (i\upsilon _l ;x,x')\frac{d}{{dT}}G(i\upsilon _l ;x,x')].
\end{equation}
According to above relation, in terms of free energy, entropy can be rewritten in the following form
\begin{equation}
S =  - \frac{E}{T} - T\frac{d}{{dT}}(\frac{E}{T}),
\end{equation}
and internal energy of the system becomes
\begin{equation}\label{U}
U =  - T^2 \frac{d}{{dT}}(\frac{E}{T}).
\end{equation} 

\section{Examples}\label{Example}
\subsection{1+1 Dimension}\label{1+1 Dimension}
\noindent In this part, we obtain the force induced, casimir entropy, and internal energy of the system which consist of two one-dimension objects with
susceptibilities $\chi_1(\omega)$ and $\chi_2(\omega)$ in the presence of fluctuating massless scalar field. In this case, the fluctuating field is defined in
$(1+1)$-dimensional space-time $(x=(\textbf{x},t)\in\mathbb{R}^{1+1})$. The Green's function of the system and
reservoir are given by
\begin{equation}\label{Green function 3}
G^0(\omega;\textbf{x}-\textbf{x}')=\frac{e^{-i\omega|\textbf{x}-\textbf{x}'|}}{2\omega},
\end{equation}
and
\begin{equation}\label{medium green}
G^0_{\omega'}(x-x')=\delta(\textbf{x}-\textbf{x}')\Theta(t-t')\frac{e^{-i{\omega}'(t-t')}}{2{\omega}'},
\end{equation}
respectively.
Using Eqs.(\ref{exp1}) and (\ref{Green function 3}), the free energy of 1+1D in the first approximation becomes
\begin{equation}\label{system lagrangian}
E =  - K_B T\sum\limits_{l = 1}^\infty  {\int {dx} \int {dx'} } \frac{{e^{ - 2\alpha _l \left| {x - x'} \right|} }}{{(2\alpha _l )^2 }}\chi (i\alpha _l ,x)\chi (i\alpha _l ,x'),
\end{equation}
where
\begin{equation}\label{a}
\alpha _l  = \frac{{\upsilon _l }}{C} = \frac{{2\pi lK_B }}{{\hbar C}}.
\end{equation}
We assume that susceptibility is position independent. The matter distribution is homogeneous in $\Omega_1$ and $\Omega_2$ intervals, so the dielectric function is defined by
\begin{equation}
\frac{{\varepsilon (\omega ,x)}}{{\varepsilon _0 }} = \left\{ {\begin{array}{*{20}c}
   {\frac{{\varepsilon _1 (\omega )}}{{\varepsilon _0 }},\,\,\,\,\,\,a < x < b}  \\
   {1,\,\,\,\,\,\,\,\,\,\,\,\,\,\,\,\,\,\,b < x < c}  \\
   {\frac{{\varepsilon _2 (\omega )}}{{\varepsilon _0 }},\,\,\,\,\,\,\,c < x < d}  \\
\end{array}} \right.
\end{equation}
 The free energy at finite temperature is given by
\begin{equation}
\begin{array}{l}
 E =  - K_{\bf B} T\sum\limits_{l = 1}^\infty  {\frac{1}{{(2\alpha _l T)^2 }}[\chi _1^2 \frac{{(b - a)}}{{\alpha _l T}} + \chi _2^2 \frac{{(d - c)}}{{\alpha _l T}}}  \\ 
 \,\,\,\,\,\,\,\,\,\,\,\,\,\,\,\,\,\,\,\,\,\,\,\,\,\,\,\, - \frac{{2\chi _1 \chi _2 }}{{(2\alpha _l T)^2 }}(e^{ - 2\alpha _l Td}  - e^{ - 2\alpha _l Tc} )(e^{2\alpha _l Tb}  - e^{2\alpha _l Ta} )], \\ 
 \end{array}
\end{equation}
which consists of self energies and interaction energies. By defining 
\begin{equation}
\begin{array}{l}
 a + b = 2r_1 ,\,\,b - a = 2r'',\,\,c + d = 2r_2  \\ 
 d - c = 2r',\,\,r = r_2  - r_1 , \\ 
 \end{array}
\end{equation}
the free energy is obtained in terms of distances 
\begin{equation}
\begin{array}{l}
 E =  - K_{\bf B} T\sum\limits_{l = 1}^\infty  {\frac{1}{{(2\alpha _l T)^2 }}[\chi _1^2 \frac{{r''}}{{\alpha _l T}} + \chi _2^2 \frac{{r'}}{{\alpha _l T}}}  \\ 
 \,\,\,\,\,\,\,\,\,\,\,\,\,\,\,\,\,\,\,\,\,\,\,\,\,\,\,\, - \frac{{2\chi _1 \chi _2 }}{{(2\alpha _l T)^2 }}e^{ - 2\alpha _l Tr} (e^{ - 2\alpha _l T(r' + r'')}  - e^{2\alpha _l T(r' + r'')}  + e^{2\alpha _l T(r' - r'')}  - e^{ - 2\alpha _l T(r' - r'')} )]. \\ 
 \end{array}
\end{equation}
According to Eq.(\ref{2}), the entropy of the system becomes
\begin{equation}
\begin{array}{l}
 S =  - K_B [ - \chi _1^2 r''\frac{{Zeta[3]}}{{(\gamma T)^3 }} - \chi _2^2 r'\frac{{Zeta[3]}}{{(\gamma T)^3 }} \\ 
 \,\,\,\,\,\,\,\,\,\,\,\,\, + \frac{{\chi _1 \chi _2 }}{{2(\gamma T)^2 }}(Li_2 (e^{ - 2\gamma T(r + r' - r'')} ) - Li_2 (e^{ - 2\gamma T(r + r' + r'')} ) - Li_2 (e^{ - 2\gamma T(r - r' - r'')} ) + Li_2 (e^{ - 2\gamma T(r - r' + r'')} )) \\ 
 \,\,\,\,\,\,\,\,\,\,\,\,\, - \frac{{\chi _1 \chi _2 }}{{\gamma T}}((r + r' - r'')Log(1 - e^{ - 2\gamma T(r + r' - r'')} ) - (r + r' + r'')Log(1 - e^{ - 2\gamma T(r + r' + r'')} ) \\ 
 \,\,\,\,\,\,\,\,\,\,\,\,\,\,\,\,\,\,\,\,\,\,\,\,\,\,\,\, - (r - r' - r'')Log(1 - e^{ - 2\gamma T(r - r' - r'')} ) + (r - r' + r'')Log(1 - e^{ - 2\gamma T(r - r' + r'')} )),\\ 
 \end{array}
\end{equation}
where $\gamma  = \frac{{2\pi K_B }}{{\hbar C}}$  and $Li_s(z)$ are polylogarithm functions which are defined by the infinite sum $Li_s (z) = \sum\limits_{k = 1}^\infty  {\frac{{z^k }}{{k^s }}} $.
In terms of temperature, the variance of entropy is shown in Fig. 1 and in Fig. 2.\\
 The force induced resulting from the fluctuating massless scalar field on two objects is given by
\begin{equation}
\begin{array}{l}
 F = \frac{{\partial E}}{{\partial r}} =  - K_B \frac{{\chi _1 \chi _2 }}{{2\gamma }}( - Log[1 - e^{ - 2\gamma T(r + r' + r'')} ] + Log[1 - e^{ - 2\gamma T(r - r' + r'')} ] \\ 
 \,\,\,\,\,\,\,\,\,\,\,\,\,\,\,\,\,\,\,\,\,\,\,\,\,\,\,\,\,\,\,\,\,\,\,\,\,\,\,\,\,\,\,\,\,\, + Log[1 - e^{ - 2\gamma T(r + r' - r'')} ] - Log[1 - e^{ - 2\gamma T(r - r' - r'')} ]). \\ 
 \end{array}
\end{equation}
Using Eq.(\ref{U}), we obtain internal energy of the system
\begin{equation}
\begin{array}{l}
 U = K_B \left[ {\frac{{3(\chi _1 ^2 r'' + \chi _2^2 r')}}{{4\gamma ^3 T^2 }} + } \right.\frac{{\chi _1 \chi _2 }}{{2\gamma ^4 T^3 }}(Li_4 (e^{ - 2\gamma T(r + r' + r'')} ) - Li_4 (e^{ - 2\gamma T(r - r' - r'')} ) \\ 
 \,\,\,\,\,\,\,\,\,\,\,\,\,\,\,\,\,\, + Li_4 (e^{ - 2\gamma T(r - r' + r'')} ) - Li_4 (e^{ - 2\gamma T(r + r' - r'')} )) \\ 
 \,\,\,\,\,\,\,\,\,\,\,\,\,\,\,\,\, - \frac{{\chi _1 \chi _2 }}{{4\gamma ^3 T^2 }}( - (r + r' + r'')Li_3 (e^{ - 2\gamma T(r + r' + r'')} ) + ( - r + r' + r'')Li_3 (e^{ - 2\gamma T(r - r' - r'')} ) \\ 
 \,\,\,\,\,\,\,\,\,\,\,\,\,\,\,\,\, + ( - r + r' - r'')Li_3 (e^{ - 2\gamma T(r - r' + r'')} ) - (r + r' - r'')Li_3 (e^{ - 2\gamma T(r - r'' + r')} )\left. ) \right]. \\ 
 \end{array}
\end{equation}

\subsection{ 2+1 dimension space-time}\label{2+1 dimension}
The behavior of two objects in the presence of the fluctuating massless scalar field, which is defined in 1+2 dimensional space- time, is investigated. Therefore, the green's function of the system becomes 
\begin{equation}\label{2+1 Green}
G_0(\omega,\textbf{x}-\textbf{x}')=\frac{i \hbar}{2\pi}K_0(i\omega|\textbf{x}-\textbf{x}'|),
\end{equation}
in which $K_0(i\omega|\textbf{x}-\textbf{x}'|)$ is the modified Bessel function of the second kind.
 Given Eq.(\ref{exp1}), in the first approximation, the free energy is obtained by
\begin{equation}\label{2+1 Free energy}
E=-\frac{K_B T}{4{\pi}^2} \sum_l ^\infty \int d^2\textbf{x} \int d^2\textbf{x}' K_0^2(\alpha_l|\textbf{x}-\textbf{x}'|)
\chi_1(i\alpha_l,\textbf{x})\chi_2(i\alpha_l,\textbf{x}'),
\end{equation}
where self energies are ignored. 
In objects whose susceptibilities are frequency independent, the casimir entropy of the system becomes
\begin{equation}
\begin{array}{l}
 S =  - \frac{{K_B }}{{4\pi ^2 }}\sum\limits_l {\int {d^2 x} \int {d^2 x'} \chi _1(x) \chi _2(x') } [K_0^2 (\gamma lT\left| {x - x'} \right|) \\ 
  - 2\gamma lT\left| {x - x'} \right|K_0 (\gamma lT\left| {x - x'} \right|)K_1 (\gamma lT\left| {x - x'} \right|)] \\ 
 \end{array}
\end{equation}
Using the asymptotic expansion, we have
\begin{equation}
\begin{array}{l}
 S =  - \frac{{K_B }}{{4\pi ^2 }}\int {d^2 x} \int {d^2 x'} \chi _1(x) \chi _2(x') [ - \frac{1}{{e^{2\gamma T\left| {x - x'} \right|}  - 1}} - \frac{{1}}{{4\gamma T\left| {x - x'} \right|}}Log[1 - e^{2\gamma T\left| {x - x'} \right|} ] \\  \\
  + \frac{1}{{16}}\frac{1}{{(2\gamma T\left| {x - x'} \right|)^2 }}{Li_2 (e^{2\gamma T\left| {x - x'} \right|} )} - \frac{1}{{8(2\gamma T\left| {x - x'} \right|)^3 }}{Li_3 (e^{2\gamma T\left| {x - x'} \right|} )}], \\ 
 \end{array}
\end{equation}
and the internal energy of the system is given by
\begin{equation}
U =  - \frac{{K_B }}{{2\pi ^2 }}\sum\limits_{l = 1}^\infty  {\int {d^2 x\int {d^2 x'} } \chi _1(x) \chi _2(x') } \alpha _l T^2 \left| {x - x'} \right| K_0 (\alpha _l T\left| {x - x'} \right|)K_1 (\alpha _l T\left| {x - x'} \right|).
\end{equation}
 
\subsection{3+1-dimensional space-time}\label{3+1-Dimansion}
\noindent In this section, we restrict ourselves to $(3+1)$-dimensional space-time $(x=(\textbf{x},t)\in\mathbb{R}^{3+1})$.
In this case, the Green's function of the system becomes
\begin{equation}\label{1+3 dimension}
G_0(\omega,\textbf{x}-\textbf{x}')=\frac{1}{4\pi}\frac{e^{-i\omega|\textbf{x}-\textbf{x}'|}}{|\textbf{x}-\textbf{x}'|},
\end{equation}
and the free energy in the first approximation is given by
\begin{equation}\label{3+1 E.F}
E=-\frac{{K_B T}}{{16\pi }}\sum\limits_{l=0}^\infty \int d^3\textbf{x} \int d^3 \textbf{x}' \frac{e^{-2\nu_l|\textbf{x}-
\textbf{x}'|}\chi_1(\textbf{x},\nu_l)\chi_2(\textbf{x}',\nu_l)}{|\textbf{x}-\textbf{x}'|^2}.
\end{equation}
When susceptibilities are independent of frequency, we have
\begin{equation}
E =  - \frac{{K_B T}}{{16\pi }}\int {d^3 x} \int {d^3 x'} \frac{{\chi _1 (x)\chi _2 (x)}}{{\left| {x - x'} \right|^2 }}(\frac{1}{{1 - e^{ - 2\gamma \left| {x - x'} \right|} }})
\end{equation}
The system's internal energy is achieved by
\begin{equation}
U = \frac{{K_B }}{{8\pi }}\int {d^3 x\int {d^3 x'} } \gamma T^2 \frac{{\chi _1 (x)\chi _2 (x')}}{{\left| {x - x'} \right|}}\frac{{e^{ - 2\gamma T\left| {x - x'} \right|} }}{{(1 - e^{ - 2\gamma T\left| {x - x'} \right|} )^2 }},
\end{equation}
and the casimir entropy is given by
\begin{equation}
S =  - \frac{{K_B }}{{16\pi }}\int {d^3 x} \int {d^3 x'} \frac{{\chi _1 (x)\chi _2 (x')}}{{\left| {x - x'} \right|^2 }}[\frac{1}{{1 - e^{ - 2\gamma T\left| {x - x'} \right|} }} + \frac{{2\gamma T\left| {x - x'} \right|e^{ - 2\gamma T\left| {x - x'} \right|} }}{{(1 - e^{ - 2\gamma T\left| {x - x'} \right|} )^2 }}].
\end{equation}
In the lower temperature, we can use the expansion form of the previous relation to obtain
\begin{equation}\label{pp}
S =  - \frac{{K_B }}{{16\pi }}\int {d^3 x} \int {d^3 x'} \frac{{\chi _1 (x)\chi _2 (x)}}{{\left| {x - x'} \right|^2 }}[\frac{1}{{\gamma T\left| {x - x'} \right|}} + \frac{1}{2} + \frac{{(\gamma T)^3 \left| {x - x'} \right|}}{{45}} - \frac{{4(\gamma T)^5 \left| {x - x'} \right|^3 }}{{945}} + ...].
\end{equation}
We consider two spheres, with radius $a$ and $b$. The distance between their centers is $R>a+b$.
The susceptibilities of the spheres are $\chi_1(\textbf{x})=\chi_1 \delta(r-a)$ and $\chi_2(\textbf{x})=\chi_2 \delta(r'-b)$,
where $r$ and $r'$ are radial coordinates in spherical coordinate systems, and $R$ lies along the $z$ axis of both coordinate
systems. The distance between points on the spheres is
\begin{eqnarray}\label{distance}
|\textbf{x}-\textbf{x}'|=\sqrt{R^2+a^2+b^2-2ab \cos\gamma-2 R(a \cos\theta-b \cos\theta')}, \nonumber\\
 \cos\gamma=\cos\theta \cos\theta'+\sin\theta \sin\theta' \cos(\phi-\phi').
\end{eqnarray}
According to Eq.(\ref{pp}), we have
\begin{equation}
\begin{array}{l}
 S =  - \frac{{K_B }}{{16\pi }}\int {d\Omega } \int {d\Omega '} \chi _1 \chi _2 [\frac{1}{{\gamma T\left| {x - x'} \right|^3 }} \\ 
  + \frac{1}{{2\left| {x - x'} \right|^2 }} + \frac{{\left( {\gamma T} \right)^3 \left| {x - x'} \right|}}{{45}} - \frac{4}{{45}}\left( {\gamma T} \right)^5 \left| {x - x'} \right|^3  + ...] \\ 
 \end{array}
\end{equation}
By using 
\begin{equation}
\int {d\Omega \int {d\Omega '} \left| {x - x'} \right|^p  = (4\pi )^2 R^p P_p (\hat a,\hat b)}, 
\end{equation}
where
\begin{equation}
\begin{array}{l}
 P_{ - 1}  = 1,\, \\ 
 P_{ - 2}  = \frac{1}{{4\hat a\hat b}}(Ln[\frac{{1 - (\hat a + \hat b)^2 }}{{1 - (\hat a - \hat b)^2 }}] + \hat aLn[\frac{{(\hat b + 1)^2  - \hat a^2 }}{{(\hat b - 1)^2  - \hat a^2 }}] + \hat bLn[\frac{{(\hat a + 1)^2  - \hat b^2 }}{{(\hat a - 1)^2  - \hat b^2 }}]) \\ 
 P_{ - 3}  =  - \frac{1}{{4\hat a\hat b}}Ln[\frac{{1 - (\hat a + \hat b)^2 }}{{1 - (\hat a - \hat b)^2 }}] \\ 
 P_p (\hat a,\hat b) = \frac{1}{{4\hat a\hat b}}\frac{1}{{(p + 2)(p + 3)}}[(1 + \hat a + \hat b)^{p + 3}  + (1 - \hat a - \hat b)^{p + 3}  \\ 
 \,\,\,\,\,\,\,\,\,\,\,\,\,\,\,\,\,\,\,\,\,\,\,\,\,\,\,\,\,\,\,\,\,\,\,\,\,\,\,\,\,\,\,\,\,\,\,\,\,\,\,\,\,\,\,\,\,\, - (1 - \hat a + \hat b)^{p + 3}  - (1 + \hat a - \hat b)^{p + 3} ] \\ 
 \,\,\,\,\,\,\,\,\,\,\,\,\,\,\,\,\,\,\,\,\,\,\,\,\,\,\,\,\,\,\,\,\,\,\,\,\,\,\,\,\,\,\,\,\,\,\,\,\,\,\,\,\,\,\,\,\,\,\,\,\,\,\,\,\,\,\,\,\,\,\,\,\,\,p = 0,1,2,3,... ,\\ 
 \end{array}
\end{equation}
where, $\hat{a}=\frac{a}{R}$ and $\hat{b}=\frac{b}{R}$
and
\begin{equation}
P_{p - 1}  = \frac{{R^{ - p} }}{{1 + p}}\frac{\partial }{{\partial R}}R^{p + 1} P_p (\hat a,\hat b),
\end{equation}
and the casimir entropy of this system is obtained by
\begin{equation}
\begin{array}{l}
 S =  - \frac{{K_B }}{{16\pi }}\chi _1 \chi _2 \left[ {\frac{1}{{\gamma T}}(4\pi )^2 R^{ - 3} P_{ - 3} (\hat a,\hat b) + \frac{1}{2}(4\pi )^2 R^{ - 2} P_{ - 2} (\hat a,\hat b)} \right. \\ 
 \,\,\,\,\,\,\,\,\,\,\,\,\,\,\,\,\,\,\,\,\,\,\,\,\,\,\,\,\,\,\,\,\, + \frac{{(\gamma T)^3 }}{{45}}(4\pi )^2 RP_1 (\hat a,\hat b) - \frac{4}{{945}}(\gamma T)^5 (4\pi )^2 R^3 P_3 (\hat a,\hat b) + \left. {...} \right] \\ 
 \end{array}
\end{equation}
The casimir entropy becomes positive in all temperatures. 

\section{Electromagnetic field}\label{ELECTROMAGNETIC FIELD}

\noindent We use the previous approach in the previous section to find the partition function in terms of the susceptibility of the medium as follows
\begin{eqnarray}\label{expansion free energy.EM}
E &=& \textit{k}_B T \sum_{l=0}^\infty \sum_{n=1}^\infty \frac{(-1)^{n+1}}{n}\int d^3\textbf{x}_1\cdots d^3
\textbf{x}_n G^0_{i_1i_2}(i\nu_l;\textbf{x}_1-\textbf{x}_2)...G^0_{i_ni_1}(i\nu_l;\textbf{x}_n-\textbf{x}_1) \nonumber \\
& \times & \chi(i\nu_l,\textbf{x}_1)\cdots\chi(i\nu_l,\textbf{x}_n),
\end{eqnarray}
where the free Green's function of the electromagnetic field is $G^0_{ij}(\textbf{x}-\textbf{x}',i\nu_l)$ . We introduce $\textbf{r}=\textbf{x}-\textbf{x}'$ and find
\begin{equation}\label{G.EM}
G_{ij}^0(\textbf{r},i\nu_l)=\frac{\nu_l^2}{c^2}\frac{e^{-\frac{\nu_lr}{c}}}{4\pi r}[\delta_{ij}(1+\frac{c}{\nu_lr}+
\frac{c^2}{\nu_l^2 r^2})-\frac{r_ir_j}{r^2}(1+\frac{3c}{\nu_l r}+\frac{3c^2}{\nu_l^2 r^2})]+\frac{1}{3}\delta_{ij}
\delta^3(\textbf{r}).
\end{equation}
We calculate the interaction energy, casimir entropy, and internal energy  of a system which
are composed of two dielectrics with volumes $V_1$ and $V_2$ and susceptibilities $\chi_1$ and $\chi_2$, respectively.
The first relevant nonzero term corresponds to $n=2$, which is given by
\begin{equation}\label{E example}
E=-\frac{1}{2}k_B T\sum_{l=0}^\infty \int_{V_1}\int_{V_2} d^3\textbf{x} d^3\textbf{x}' G^0_{ij}(\textbf{x}-
\textbf{x}',i\nu_l)G^0_{ji}(\textbf{x}'-\textbf{x},i\nu_l)
\chi_1(i\nu_l,\textbf{x})\chi_2(i\nu_l,\textbf{x}').
\end{equation}
Substituting the Green's function (\ref{G.EM}) for (\ref{E example}), we find
\begin{equation}\label{E.EM}
E=-k_B T\sum_{l=0}^\infty \int_{V_1}\int_{V_2} d^3\textbf{x} d^3\textbf{x}' \chi_1(i\nu_l,\textbf{x})
\chi_2(i\nu_l,\textbf{x}')h(\nu_l,|\textbf{x}-\textbf{x}'|),
\end{equation}
where we define
\begin{equation}\label{h}
h(\nu_l,|\textbf{x}-\textbf{x}'|)=\frac{e^{-\frac{2\nu_l}{c}|\textbf{x}-\textbf{x}'|}}{8{\pi}^2}
\{\frac{(\frac{\nu_l}{c})^4}{|\textbf{x}-\textbf{x}'|^2}
+\frac{2(\frac{\nu_l}{c})^3}{|\textbf{x}-\textbf{x}'|^3}+\frac{5(\frac{\nu_l}{c})^2}{|\textbf{x}-\textbf{x}'|^4}
+\frac{6\frac{\nu_l}{c}}{|\textbf{x}-\textbf{x}'|^5}+\frac{3}{|\textbf{x}-\textbf{x}'|^6} \}.
\end{equation}

Whenever the susceptibilities are independent of frequency, 
 we do the summation over $l$ and use the expansion of exponential term as follows
\begin{equation}
\begin{array}{l}
 E =  - K_B T\chi _1 \chi _2 \int {dx^3 } \int {dx'^3} [\frac{{55}}{{\gamma T\left| {x - x'} \right|^7 }} + \frac{3}{{2\left| {x - x'} \right|^6 }} - \frac{{\gamma T}}{{4\left| {x - x'} \right|^5 }} \\ 
 \\
 \,\,\,\,\,\,\,\,\,\,\,\,\,\,\,\,\,\,\,\,\, - \frac{{\gamma ^3 T^3 }}{{240\left| {x - x'} \right|^3 }} + \frac{{73\gamma ^5 T^5 }}{{30240\left| {x - x'} \right|}} - \frac{{197\gamma ^7 T^7 }}{{352800}}\left| {x - x'} \right| + ...]. \\ 
 \end{array}
\end{equation}
Therefore, the internal energy of the system is obtained by
\begin{equation}
\begin{array}{l}
 U = K_B \chi _1 \chi _2 \int {d^3 x} \int {d^3 x'} [\frac{{55}}{{\gamma \left| {x - x'} \right|^7 }} + \frac{\gamma }{{4\left| {x - x'} \right|^5 }}T^2  \\ 
 \\
 \,\,\,\,\,\,\,\,\,\,\, + \frac{{\gamma ^3 }}{{80\left| {x - x'} \right|^3 }}T^4  - \frac{{73\gamma ^5 }}{{6048\left| {x - x'} \right|}}T^6  + ....]. \\ 
 \end{array}
\end{equation}
The casimir entropy of the system is given by
\begin{equation}
\begin{array}{l}
 S =  - K_B \chi _1 \chi _2 \int {dx^3 } \int {dx'^3} [\frac{3}{{2\left| {x - x'} \right|^6 }} - \frac{{\gamma T}}{{2\left| {x - x'} \right|^5 }} \\ 
 \\
 \,\,\,\,\,\,\,\,\,\,\,\,\,\,\,\,\,\,\,\,\, - \frac{{\gamma ^3 T^3 }}{{60\left| {x - x'} \right|^3 }} + \frac{{73\gamma ^5 T^5 }}{{5040\left| {x - x'} \right|}} - \frac{{197\gamma ^7 T^7 }}{{50400}}\left| {x - x'} \right| + ...]. \\ 
 \end{array}
\end{equation}
For two spheres of radii a and b, the distance between their centers $R>a+b$, with susceptibilities $\chi_1=\chi_1\delta(r-a)$ and $\chi_2=\chi_2\delta(r'-a)$ , where r and $r'$ are radial coordinates in spherical  coordinate systems, Casimir entropy in natural unit becomes
\begin{equation}
\begin{array}{l}
 S =  - \chi _1 \chi _2 (4\pi )^2 [\frac{3}{2}R^{ - 6} P_{ - 6} (\hat a,\hat b) - \pi R^{ - 5} P_{ - 5} (\hat a,\hat b)T - \frac{2}{{15}}\pi ^3 R^{ - 3} P_{ - 3} (\hat a,\hat b)T^3  \\ 
 \,\,\,\,\,\,\,\,\,\,\,\,\,\,\,\,\,\,\,\,\,\,\,\,\,\,\,\, + \frac{2}{{315}}\pi ^5 R^{ - 1} P_{ - 1} (\hat a,\hat b)T^5  - 0.5\pi ^7 RP_1 (\hat a,\hat b)T^7  + ...]. \\ 
 \end{array}
\end{equation}
Casimir entropy of the system related to temperature is shown in fig.4, in which entropy becomes negative in small interval of temperature variations.\\ 
\section{Conclusion}
  \noindent In this paper, casimir entropy and internal energy of arbitrary shaped objects immersed in the fluctuating massless scalar field and the electromagnetic field are calculated by path integral methodes. The casimir entropy of the objects which are immersed in the massless scalar field in all temperatures becomes positive. However, the value of entropy observed in small interval of temperature variations of two nanospheres in the presence of electromagnetic field is negative. The result is in agreement with multiple scattering methods. 
  
\section{َAppendices}
\subsection{ Massless Scalar Field}\label{SCALAR FIELD}
\noindent The Lagrangian of total system consists of a massless scalar field in
$N+1$-dimensional space-time $(x=(\textbf{x},t)\in
\mathbb{R}^{N+1})$. Lagrangian of the scalar field, medium and interaction between scalar field and medium, respectively, are given by
\begin{equation}\label{system lagrangian}
{\cal L}_s= \frac{1}{2}\partial_\mu \varphi(x) \partial^ \mu
\varphi(x).
\end{equation}
\begin{equation}\label{reservior lagrangian}
{\cal L}_{res}= \frac{1}{2}\int\limits _0 ^\infty d\omega \left(\dot{Y}_{\omega} ^2 (x)-
{\omega}^2 Y_{\omega}^2 (x) \right).
\end{equation}
\begin{equation}\label{interaction lagrangian}
{\cal L}_{int}=\int \limits_0 ^\infty d\omega f(\omega,\textbf{x})\dot{Y}_\omega (x)\varphi(x),
\end{equation}
where $f(\omega,\textbf{x})$ is the coupling of the
scalar field with its medium. When we have the total Lagrangian, we can
quantize the total system using path-integral techniques
. Generating functional is an important quantity in any field theory with n-point correlation functions, which is obtained by taking successive functional derivatives. Here, two-point correlation functions in terms of the susceptibility of the medium is calculated \cite{kheirandish1,greiner1}.
The interacting generating-function becomes
\begin{eqnarray}\label{generating function}
W[J,J_\omega]&=&e^{\frac{i}{\hbar} \int d^n\textbf{x} \int dt \int \limits_0 ^\infty d\omega f(\omega,\textbf{x})
 (\frac{\hbar}{i} \frac{\delta}{\delta J(x)}) \frac{\partial}{\partial t} (\frac{\hbar}{i} \frac{\delta}
 {\delta J_\omega (x)})} W_0[J,J_\omega] \nonumber \\
&=&N e^{-i\hbar \int d^n\textbf{x} \int dt \int \limits_0 ^\infty d\omega f(\omega,\textbf{x}) \frac{\delta}{\delta J(x)}
\frac{\partial}{\partial t} \frac{\delta}{\delta J_\omega (x)}} \nonumber \\
 &\times & e^{-\frac{1}{2 \hbar^2} \int d^n\textbf{x} \int dt \int d^n\textbf{x}' \int dt' J(x)G^0(x-x')J(x')} \nonumber \\
 &\times& e^{-\frac{1}{2 \hbar^2} \int d^n\textbf{x} \int dt \int d^n\textbf{x}' \int dt' \int \limits _0 ^\infty
 d\omega J_\omega(x)G_\omega^0(x-x')J_\omega(x')},
\end{eqnarray}
where
\begin{eqnarray}\label{Green functions}
G^0 (x-x') &= & i\hbar \int {\frac{d^n\textbf{k} d k_0}{(2 \pi)^{n+1}}} {\frac {e^{-i\textsl{k}(\textbf{x}-\textbf{x}')}
e^{ik_0(t-t')}}{k_0^2-\textbf{k}^2+i\epsilon}},\hspace{1cm}(\epsilon>0), \nonumber \\
G_\omega ^0 (x-x') &=& i\hbar \delta^n(\textbf{x}-\textbf{x}')
\int {\frac{d k_0}{2 \pi}} {\frac {e^{ik_0(t-t')}}{(k_0)^2- \omega
^2 +i\epsilon}},
\end{eqnarray}
The two-point function can be obtained as
\begin{equation}\label{2 point Green function}
G(x-x')=(\frac{\hbar}{i})^2 \frac{\delta ^2}{\delta J(x) \delta J(x')} W[J,J_\omega]\mid_{j,j_\omega=0}.
\end{equation}
Using Eq.(\ref{generating function}), we find the following expansion of
Green's function in frequency variable
\begin{eqnarray}\label{expansion Green function}
G(\textbf{x}-\textbf{x}',\omega)=G^0(\textbf{x}-\textbf{x}',\omega)+ \int_\Omega d^n \textbf{z}_1 G^0(x-z_1,\omega)
[\omega^2 \tilde{\chi}(\omega,\textbf{z}_1)]G^0(\textbf{z}_1-\textbf{x}',\omega)+ \nonumber\\
\int_\Omega \int_\Omega d^n \textbf{z}_1 d^n \textbf{z}_2
G^0(\textbf{x}-\textbf{z}_1,\omega) [\omega^2
\tilde{\chi}(\omega,\textbf{z}_1)]G^0(\textbf{z}_1-\textbf{z}_2,\omega)
[\omega^2
\tilde{\chi}(\omega,\textbf{z}_2)]G^0(\textbf{z}_2-\textbf{x}',\omega)
+\cdots,
\end{eqnarray}
in which $\tilde{\chi}(\omega,\textbf{x})$ is the susceptibility
function of the medium in frequency variable.
The partition function of a real scalar field in the presence of a medium becomes
\begin{equation}\label{partition functions}
\Xi=\int D\varphi e^{\frac{i}{\hbar}S}=\int D\varphi e^{\frac{i}{\hbar} \int d^{n+1}x  {\cal L}},
\end{equation}
where ${\cal{L}}$ is given by Eq.(\ref{system lagrangian}). The partition function can also be written in frequency
\cite{kapusta}.
 Free energy is given in $ E=-\frac{\hbar}{\tau} \ln \Xi$, where $\tau$ is the interaction duration which becomes quite large.\\
  Using standard path-integral techniques, we find the free energy at finite temperature $T$
\begin{equation}\label{free energy}
E=\textit{k}_B T \sum _{l=0} ^\infty \ln \det[\hat{K}(\nu_l;\textbf{x},\textbf{x}')],
\end{equation}
where $\nu_l=2 \pi l \textit{k}_B T/\hbar$ are Matsubara
frequencies, $\textit{k}_B$ is Boltzman constant, and $\hat{K}(\nu_l;\textbf{x},\textbf{x}')=G^{-1}(i
\nu_l;\textbf{x},\textbf{x}')$. Therefore, we have
\begin{equation}\label{free energy 2}
E=-\textit{k}_B T \sum _{l=0} ^\infty tr \ln[G(i\nu_l;\textbf{x},\textbf{x}')].
\end{equation}
\subsection{ Electromagnetic field}\label{electromagnetic field}
\noindent The casimir energy of the medium in the presence of electromagnetic is obtained from two-point Green's function. For this purpose, the total Lagrangian density can be written in coulomb gauge $\nabla\cdot\textbf{A}=0, \, A^0=0$, as follows

\begin{equation}\label{1}
{\cal{L}}=\frac{1}{2}(\textbf{E}^2-\textbf{B}^2)+\frac{1}{2}\int_0^\infty d\omega (\dot{\textbf{Y}}_\omega ^2(x)-\omega^2 \textbf{Y}_\omega ^2(x))+\int d\omega f(\omega,\textbf{x})\textbf{A}.\dot{\textbf{Y}}_\omega.
\end{equation}
The interacting generating functional is given by
\begin{eqnarray}\label{1.W,EM}
W&=&\int D[\textbf{A}]\prod_\omega D[\textbf{Y}_\omega]\exp\frac{i}{\hbar}\int d^4x [-\frac{1}{2}A_\mu
\hat{K}_{ij}A_j-\int_0^\infty d\omega \frac{1}{2} Y_{\omega,i}(\partial_t^2+\omega^2)\delta_{ij}Y_{\omega,j} \nonumber\\
&+&\int _0^\infty d\omega f(\omega,\textbf{x})A_i\dot{Y}_{\omega,i}+ J_i A_i +\int_0^\infty d\omega J_{\omega,i}
Y_{\omega,i}],
\end{eqnarray}
where the kernel $\hat{K}_{ij}$ is defined by
\begin{equation}\label{kernel}
\hat{K}_{ij}=(\frac{\partial_0^2}{c^2}-\nabla^2)\delta_{i,j}-\partial_i \partial_j.
\end{equation}
Using the well known relation
\begin{equation}\label{G}
G_{ij}(x,x')=(\frac{\hbar}{i})^2\frac{\delta^2}{\delta J_i(x)\delta J_j(x')} W[j,{j_\omega}]|_{j,{j_\omega}=0},
\end{equation}
and following the same process that we did for the scalar case, we obtain the following expansion for Green's function
\begin{eqnarray}\label{expansion Green function,EM}
G_{ij}(\textbf{x}-\textbf{x}',\omega)=G^0_{ij}(\textbf{x}-\textbf{x}',\omega)+ \int_\Omega d^3 \textbf{z}_1 G^0_{il}
(\textbf{x}-\textbf{z}_1,\omega)[\omega^2 \tilde{\chi}(\omega,\textbf{z}_1)]G^0_{lj}(\textbf{z}_1-\textbf{x}',\omega)+
\nonumber\\
\int_\Omega \int_\Omega d^3 \textbf{z}_1 d^3 \textbf{z}_2 G^0_{il}(\textbf{x}-\textbf{z}_1,\omega)[\omega^2
\tilde{\chi}(\omega,\textbf{z}_1)]G^0_{lm}(\textbf{z}_1-\textbf{z}_2,\omega)[\omega^2
\tilde{\chi}(\omega,\textbf{z}_2)]G^0_{mj}(\textbf{z}_2-\textbf{x}',\omega) +\cdots,\nonumber\\
\end{eqnarray}
where $\frac{\epsilon(\omega)}{\epsilon_0}=\chi(\omega)+1$.

\bibliographystyle{apsrev4-1}

\newpage
\begin{figure}
\begin{center}
\includegraphics[width=3.74in, height=2.74in]{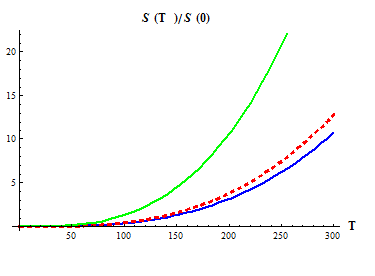}
\end{center}
 \textbf{fig.1} Entropy of two nano ribbons in one dimension with the same susceptibility in terms of temperature, blue(b-a=2nm,c-b=8nm, d-c=4nm), red(b-a=2nm,c-b=8nm,d-c=8nm),green(b-a=10nm,c-b=8nm,d-c=8nm).\\
\end{figure}

\begin{figure}
\begin{center}
\includegraphics[width=3.74in, height=2.74in]{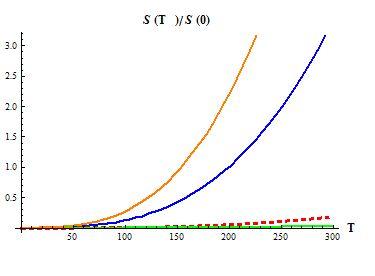}
\end{center}
\textbf{fig.2} Entropy of two nano ribbons in one dimension with (b-a=1nm, c-b=4nm, d-c=1nm) and susceptibilities blue ($\chi_1=11.68$ and $\chi_2=2.6$), red($\chi_1=11.68$ and $\chi_2=1000$), green($\chi_1=11.68$ and $\chi_2=6000$), orange($\chi_1=2$ and $\chi_2=3$) in terms of temperature.\\
\end{figure}

\begin{figure}
\begin{center}
\includegraphics[width=3.74in, height=2.74in]{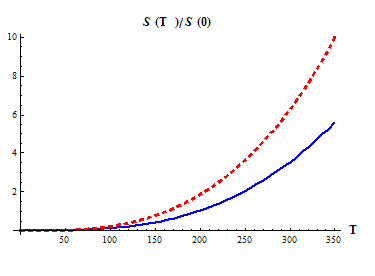}
\end{center}
\textbf{fig.3} Entropy of the two spheres of radius a and b and distance between their centers R, immersed in scalar 3+1 dimension field in terms of temperature, blue($a=1nm,b=2nm,R=10nm$) ,red($a=1nm, b=2nm, R=20nm$) and susceptibilities ($\chi_1=11.68 $ and $\chi_2=2.6$).\\
\end{figure}

\begin{figure}
\begin{center}
\includegraphics[width=3.74in, height=2.74in]{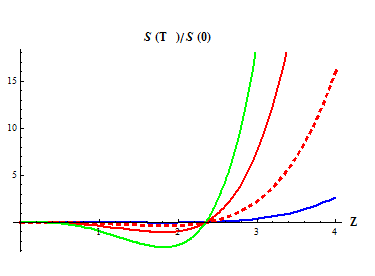}
\end{center}
\textbf{fig.4} Entropy of the two spheres, of radius a and b and distance between their centers R ($a=1nm,b=2nm,R=10mm$), immersed in electromagnetic field in terms of $Z=4\pi RT$  ,with susceptibilities, blue($\chi_1\chi_2=1$), red dash ($\chi_1\chi_2=6$), red($\chi_1\chi_2=20$), 
green($\chi_1\chi_2=50$) 
\end{figure}

\end{document}